# Study of Uniaxial Tensile Properties of Hexagonal Boron Nitride Nanoribbons

Ratul Paul, Tamanna Tasnim, Rajkumar Dhar, Satyajit Mojumder, Sourav Saha, Mohammad Abdul Motalab*
Department of Mechanical Engineering
Bangladesh University of Engineering and Technology, Dhaka-1000
*Email: mtipuz@yahoo.com, Telephone: +8801779198595, +880-2-9665636, 9665650 ext:7230

*Abstract*— **Uniaxial tensile properties of hexagonal boron nitride nanoribbons and dependence of these properties on temperature, strain rate, and the inclusion of vacancy defects have been explored with molecular dynamics simulations using Tersoff potential. The ultimate tensile strength of pristine hexagonal boron nitride nanoribbon of 26 nm x 5 nm with armchair chirality is found to be 100.5 GPa. The ultimate tensile strength and strain have been found decreasing with increasing the temperature while an opposite trend has been observed for increasing the strain rate. Furthermore, the vacancy defects reduce ultimate tensile strength and strain where the effect of bi-vacancy is clearly dominating over point vacancy.**

## I. Introduction

The hexagonal boron nitride nanostructures have received research interest due to their supreme electrical, chemical and mechanical properties. In some aspects, hBN holds advantages over its analogue graphene like electrical insulation with large band gap, [1] thermal and chemical stabilities, high thermal conductivities [2], [3] and comparable mechanical properties. [4] These remarkable properties have made hBN nanostructures to have many potential applications in engineering fields. [5]

We have studied hexagonal boron nitride nanoribbons (hBNNRs) in this paper as it is less explored than other nanostructures. As hBNNRs possess high tensile strength, [6], [7] it is possible to enhance tensile strength of polymer matrix by using them for structural reinforcement. [8], [9]. The advantage of using nanoribbons over other nanostructures lies in its active chemical edges. These active edges create strong chemical bonding between filler and matrix composites. Moreover, it is also found that mechanical properties of BNNRs can be improved by narrowing width. [10] Thus hBNNRs poses prospect in reinforcement applications of matrix composites.

Fabrication process of hBNNRs is still challenging. One way to accomplish this is by unfolding hexagonal Boron Nitride nanotubes (hBNNTs) [11]. Chemical vapor deposition (CVD) technique also shows potential in hBNNRs synthesis [12]. As seen in the case of other nanostructures, CVD incurs unavoidable defects in nanostructures. These defects impose impacts on electrical and mechanical properties of nanomaterials. [13]–[15] Sinitsii et al. [16] demonstrated potassium induced splitting of BNNTs to fabricate BNNRs and there the presence of defects was also evident. Therefore, taking into account of the inevitability of structural defects, we have studied the effect of various defects on the properties of hBNNRs.

Defects in the form of vacancies like point vacancy and edge vacancy have been presented in this paper. Mortazavi *et al.* [17] and Khan *et al.* [18] showed the effects of randomly distributed defects on the thermal conductivity and tensile response of single layer graphene sheets. Using molecular dynamics simulation with Tersoff potential, we have analyzed the change in tensile strength of hBNNRs with the inclusion of vacancies. It is observed that with the increasing concentration of vacancies, tensile strength decreases. We have also shown the effect of temperature and strain rate on the tensile strength of hBNNRs and the modified properties have been compared to that of pristine hBNNRs.

## II. Simulation Model

In this study, molecular dynamics simulations have been carried out using LAMMPS [19] (Large-scale Atomic/Molecular Massively Parallel Simulator) for calculating the tensile properties of hBNNR. To define the interatomic interactions between boron and nitrogen atoms in the molecular dynamics simulations, we used Tersoff potential with parameters obtained by Lindsay and Broido. [20] For our study, we used a hBNNR structure of 26 nm x 5 nm with armchair chirality containing 4800 atoms. In the modeling of tensile strength, we fix four rows of atoms at the two ends of the hBNNR in longitudinal direction in which uniaxial strain has been applied to prevent atoms from sublimating [21] as shown in Fig. 1.

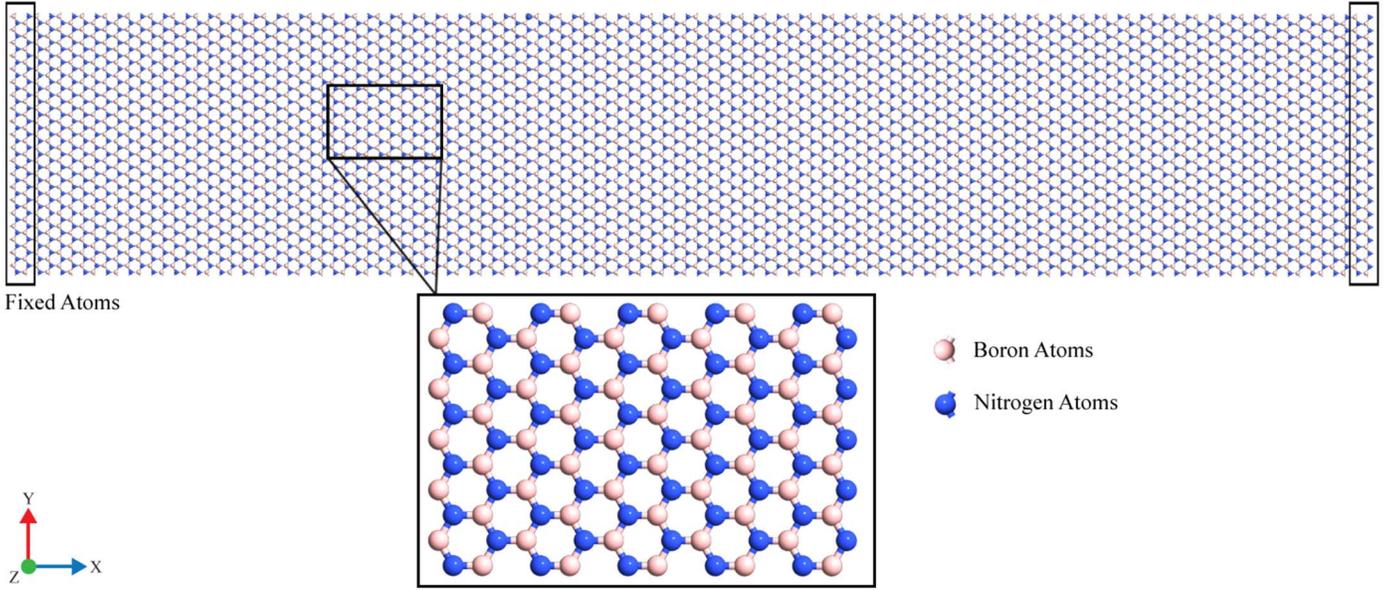

Fig. 1: Structure of 26 nm x 5 nm hBNNR with magnified view of 5 x 5 cell. Light pink atoms represent boron where blue atoms represent nitrogen.

Point vacancy was introduced in the structure by the removal of a randomly chosen atom from the geometry while the bi-vacancy was created by introducing two consecutive point vacancies or by simply removing a pair of bonding atoms [22], [23]. The defects are illustrated in Fig. 2. The thickness of multilayer hBNNR was assumed to be 0.33 nm. [11], [21], [24] Hence periodic boundary conditions have been applied along the X, Y and Z axis keeping a 4 nm gap in the simulation box of the hBNNR along the Z axis during the simulation process to ensure no interactions take place between the hBNNR in the simulation domain and its image. [23]

The equations of motion were solved using the Velocity-Verlet algorithm [25] with a time step of 0.001 ps. The initial structure was first relaxed by conjugate gradient algorithm and then equilibrated using constant pressure and constant pressure ensemble (NPT) by means of Nose´–Hoover barostat and thermostat method for 50 ps. Strain has been applied in the structure along the x direction while keeping the Y and Z boundaries in zero stress using NPT ensemble. We varied the strain applied to the x direction and virial stress were calculated at each strain level to obtain the stress-strain response of the structure using the following formula.

$$\sigma(r) = \frac{1}{V}\sum_i [(-m_i \mathring{u}_i \otimes \mathring{u}_i + \frac{1}{2}\sum_{j \neq i} r_{ij} \otimes f_{ij}$$

where, σ is the virial stress, V is the volume of the stanene atom, the summation is over all the atoms occupying the total volume, $m_i$ is the mass of atom $\mathring{u}_i$ is the time derivative which indicates the displacement of atom with respect to a reference position, $r_{ij}$ is the position vector of atom, $\otimes$ is the cross product, and $f_{ij}$ is the interatomic force applied on atom $i$ by atom $j$.

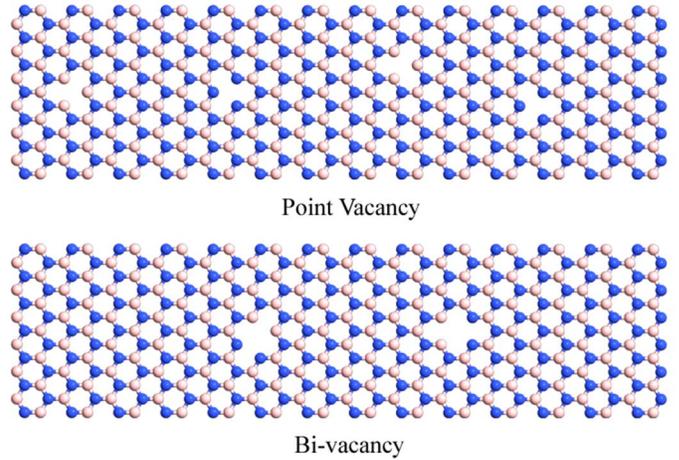

Fig. 2: Point vacancy and Bi-vacancy in armchair hBNNR structure. A total number of four atoms are absent in both structures.

## III. SIMULATION RESULTS

The stress- strain curves recorded for pristine 26 nm x 5 nm armchair hBNNR at 100K, 200K, 300K, 400K and 500K are shown in Fig. 3. All the simulations were conducted at a constant strain rate of $1\times10^9$ s$^{-1}$. It is clear that the ultimate uniaxial tensile strength and strain decrease gradually with rising temperature.

Ultimate uniaxial tensile strength drops from 106.87 GPa at 100K to 97.24 GPa at 500K and strain decreases from 0.33825 at 100K to 0.31125 at 500K. Similar trend of ultimate strength and strain had been reported for single layer hexagonal boron nitride sheets [26] and graphene sheet. [27], [28]

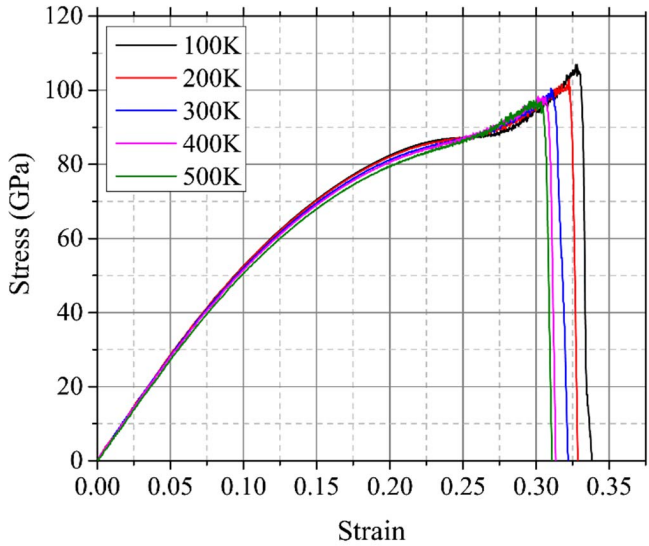

Fig. 3: Stress-strain curves of 26 nm x 5 nm armchair hBNNR at different temperature at a constant strain rate of $1\times10^9$ s$^{-1}$

The ultimate uniaxial tensile strength and strain relations with temperature for pristine hBNNR are shown in Fig. 4 and Fig. 5.

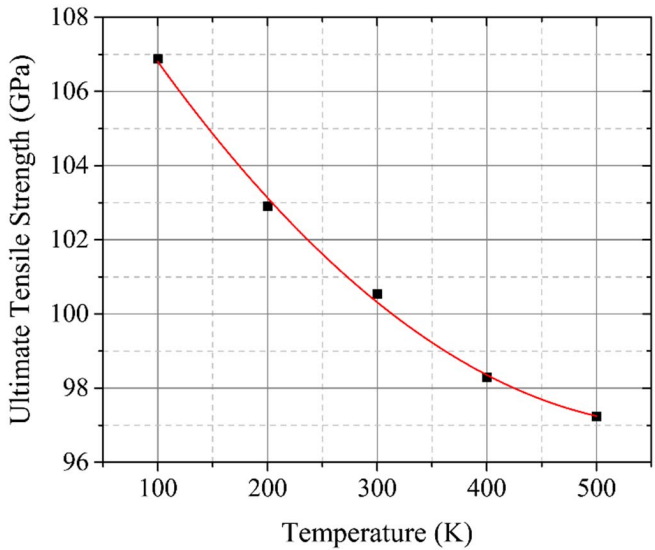

Fig. 4: Change of ultimate uniaxial tensile strength of 26 nm x 5 nm armchair hBNNR with temperature

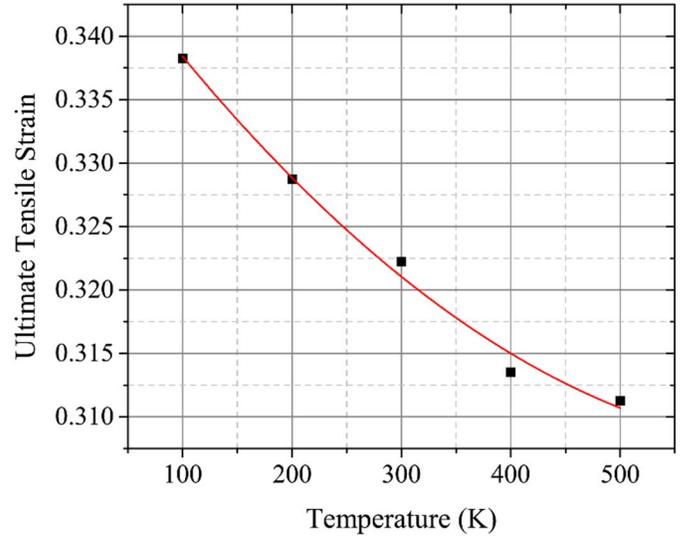

Fig. 5: Change of ultimate uniaxial tensile strain of 26 nm x 5 nm armchair hBNNR with temperature

It is well known fact that change in strain rate effects the mechanical behavior of a material and hBNNR should be no exception. Change in stress-strain behavior is seen for change in strain rate. Increase in strain rate results in increase in ultimate uniaxial tensile strength and strain which is represented in Fig. 6.

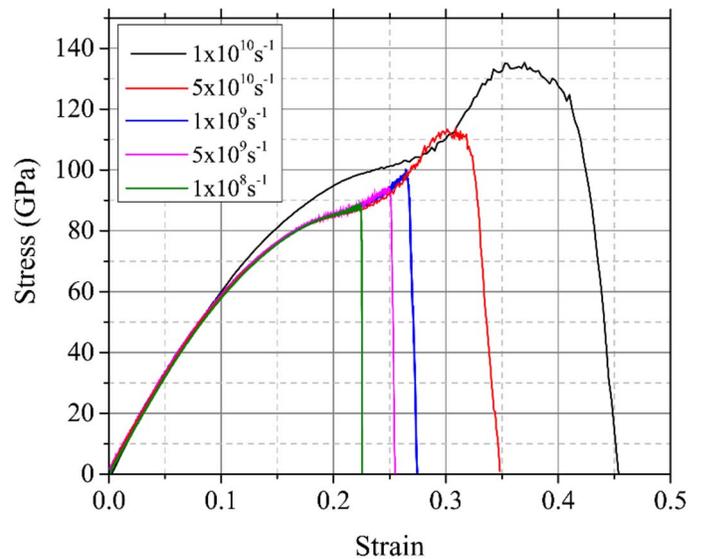

Fig. 6: Stress-strain curves of 26 nm x 5 nm armchair hBNNR at different strain rates at a constant temperature of 300K

From Fig. 7 and Fig. 8, we can see that the ultimate uniaxial tensile strength and strain rises with increase in strain rate. 34.3% increase in ultimate uniaxial tensile strength and 42.8% increase in ultimate uniaxial tensile strain is observed for increase of strain rate from $1 \times 10^8$ s$^{-1}$ to $1 \times 10^{10}$ s$^{-1}$.

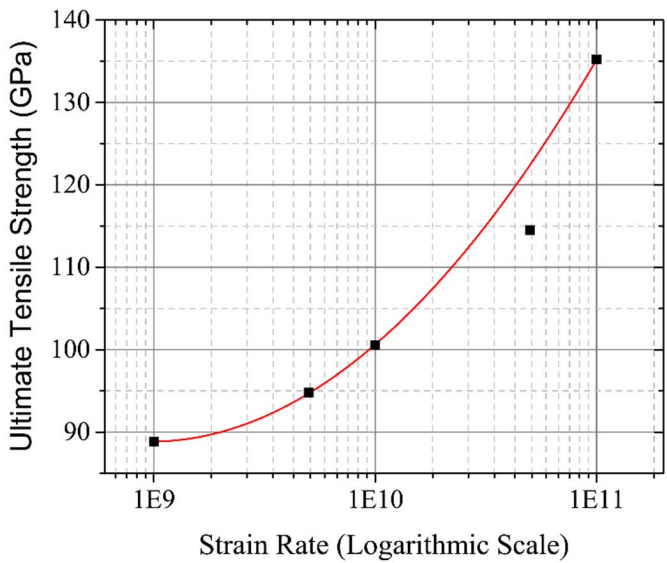

Fig. 7: Change of ultimate uniaxial tensile strength of 26 nm x 5 nm armchair hBNNR with strain rate

Similar trend in change of tensile properties for change in strain rate can be seen in previous study done on boron nitride sheets. [27]

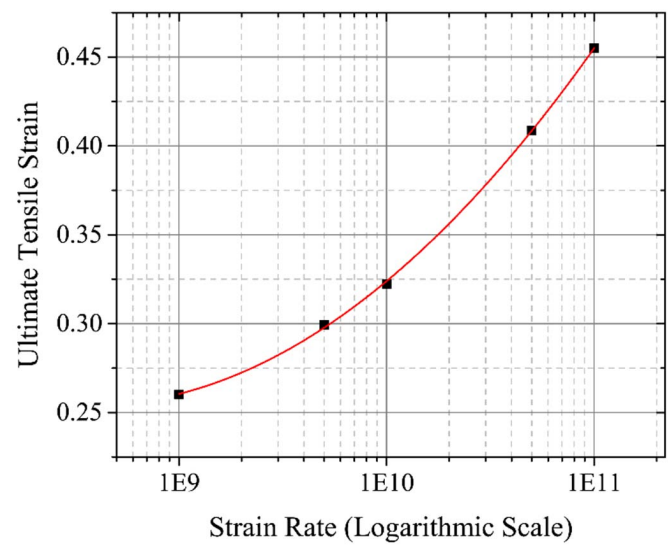

Fig. 8: Change of ultimate uniaxial tensile strain of 26 nm x 5 nm armchair hBNNR with strain rate

Stress-strain curve of hBNNR structures with point vacancy and bi-vacancy has been shown in Fig. 9 and Fig. 10.

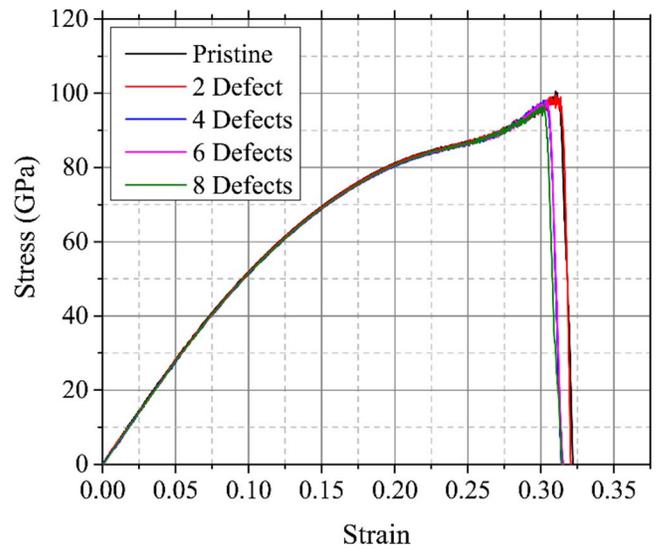

Fig. 9: Stress-strain curves of 26 nm x 5 nm armchair hBNNR with different amount of point vacancy

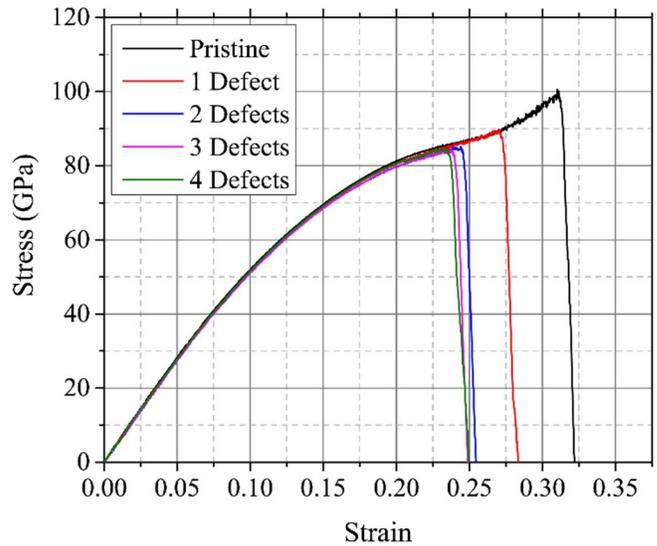

Fig. 10: Stress-strain curves of 26 nm x 5 nm armchair hBNNR with different amount of bi-vacancy

Impact of point vacancy and bi-vacancy on tensile properties of hBNNR is shown in Fig. 11 and Fig. 12. Increasing defect causes decrease in ultimate uniaxial tensile strength and strain in hBNNR. We introduced these two types of defects keeping the number of atoms removed from the structure constant. Therefore, tensile simulation results acquired form the structure with two point vacancies were compared with the one with one bi-vacancy.

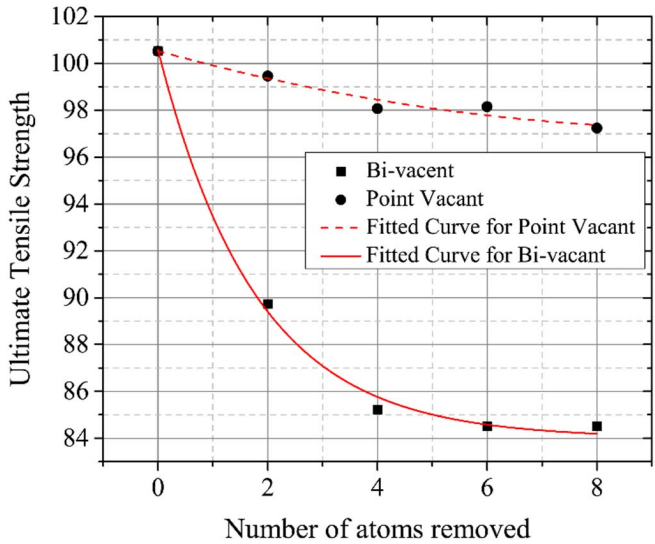

Fig. 11: Change of ultimate uniaxial tensile strength of 26 nm x 5 nm armchair hBNNR with inclusion of vacancy defects

The effect of bi-vacancy if more prominent on the tensile properties of hBNNR compared to the effects of point vacancy. There is only 3.2% decrease in ultimate uniaxial tensile strength for the removal of eight atoms as point vacancy where the decrease is 15.9% for removal of same number of atoms as bi-vacancy. Ultimate uniaxial tensile strain for the removal of similar number of atoms decreased 2.5% for point vacancy and 22.4% for bi-vacancy.

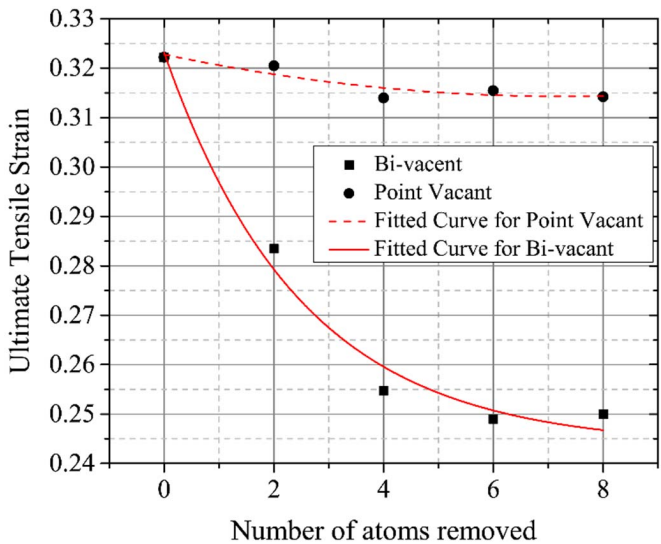

Fig. 12: Change of ultimate uniaxial tensile strain of 26 nm x 5 nm armchair hBNNR with inclusion of vacancy defects

Fracture patterns during the tensile test of pristine hBNNR and defected hBNNR are presented in Fig. 13. The fracture of hBNNR with bi-vacancy clearly shows the initiation and propagation of fracture from the defect. But although structure with point vacancy shows decreased tensile strength and strain, the fracture did not start at any vacancy. Hence, the effect of bi-vacancies is greater on the tensile properties of hBNNR than point vacancies.

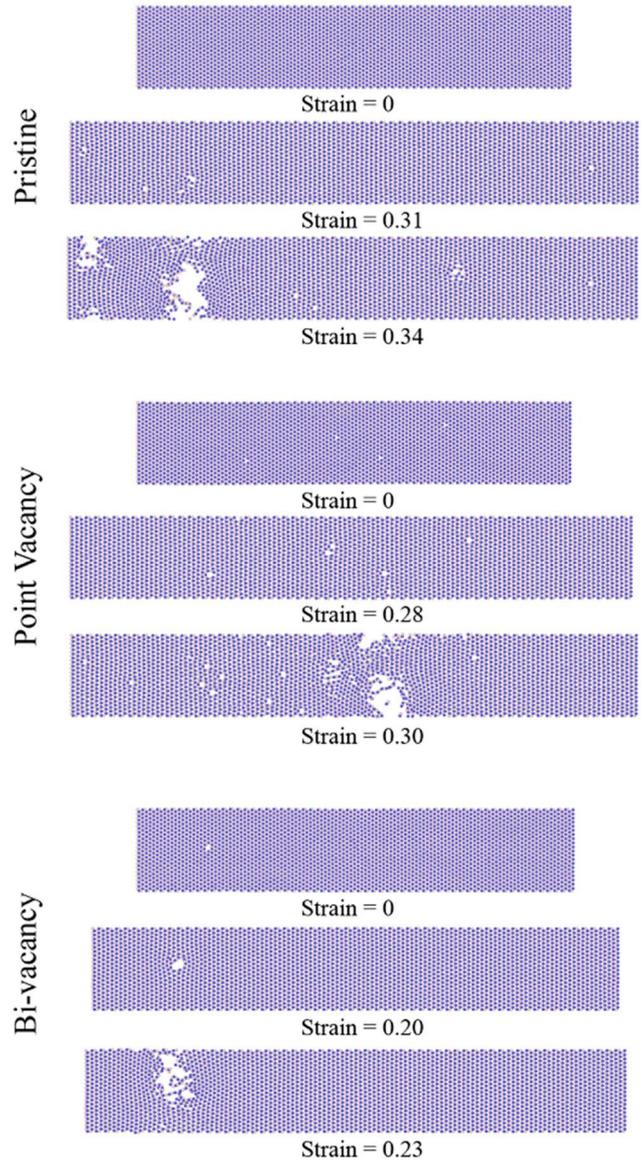

Fig. 13: Fracture patterns of hBNNR in pristine condition, with point vacancy and with bi-vacancy when uniaxial tensile strain is applied along horizontal direction (X axis)

IV. CONCLUSION

To conclude, the uniaxial tensile properties have been explored for pristine hBNNR at different temperatures, strain rates and after inclusion of defects. It is found that, ultimate tensile strength and ultimate tensile strain decrease with increasing temperature and increase with increasing strain rate.

Ultimate tensile strength and strain decreased after we introduced vacancy defects in the hBNNR structure. The effect of bi-vacancy is observed to be more prominent on the tensile properties of hBNNR compared to point vacancy.